\documentclass[aps,prl,twocolumn,floatfix, showpacs]{revtex4}
\usepackage{graphicx}
\usepackage{dcolumn}
\usepackage{bm}

\begin{document}

\title{Enhanced heat transport by turbulent two-phase Rayleigh-B\'enard convection}
\author{Jin-Qiang Zhong$^1$}
\author{Denis Funfschilling$^2$}
\author{Guenter Ahlers$^1$}
\affiliation{$^1$Department of Physics and iQCD, University of California, Santa Barbara, CA 93106, USA\\
$^2$LSGC CNRS - GROUPE ENSIC, BP 451, 54001 Nancy Cedex, France}
\date{\today}
 
\begin{abstract}
We report measurements of turbulent heat-transport in samples of ethane (C$_2$H$_6$) heated from below while the applied temperature difference $\Delta T$ straddled the liquid-vapor co-existance curve $T_\phi(P)$. When the sample top temperature $T_t$ decreased below $T_\phi$, droplet condensation occurred and the latent heat of vaporization $H$ provided an additional heat-transport mechanism.The effective conductivity $\lambda_{eff}$ increased linearly with decreasing $T_t$, and reached a maximum value $\lambda_{eff}^*$ that was an order of magnitude larger than the single-phase $\lambda_{eff}$.  As $P$  approached the critical pressure, $\lambda_{eff}^*$ increased dramatically even though $H$ vanished. We attribute this phenomenon to an enhanced droplet-nucleation rate as the critical point is approached.

\end{abstract}

\pacs{47.27.te, 47.20.Bp, 47.55.pb, 47.55.db}

\maketitle

Turbulent heat transport in a fluid heated from below  (Rayleigh-B\'enard convection or RBC) has been a topic of intense fundamental research for some  time \cite{Si94,AGL09}. Most of this work was done under conditions where the fluid was in a single phase, that is far from any thermodynamic phase-transitions. Here we focus on the case where the applied temperature difference $\Delta T = T_b - T_t$  ($T_b$ and $T_t$ are the bottom and top temperatures respectively) spans a liquid-vapor phase-transition line $T_\phi(P)$.  In that case condensation and vaporization (``boiling") can provide heat-transport mechanisms in addition to the usual turbulently advected heat current. For these mechanisms the latent heat of vaporization $H$ clearly plays a major role. This case is of great practical importance because the exceptionally high effective thermal conductivity $\lambda_{eff}$ enables numerous industrial applications, including for instance  miniaturized heat exchangers and performance enhancement in process industry.  It has been studied extensively from an engineering  viewpoint, and a wealth of empirical correlations based on these studies has been used in designs that range from miniaturized devices for cooling of computer components to large-scale power plants. \cite{Co72,Dh91,TT97,Dh98}

Here we report on a study that was intended to address some of the fundamental physical aspects of this problem. We measured $\lambda_{eff}$ in cylindrical samples of diameters $D$  about equal to their heights $L$ using ethane (C$_2$H$_6$) below but near its critical point (CP). The sample pressure $P$ and $\Delta T$ were held constant while the mean temperature $T_m = (T_t + T_b)/2$  was changed in steps much smaller than $\Delta T$ through the two-phase region. We focused primarily on the parameter range where the heat-transport enhancement $\delta \lambda_{eff}$ was due to condensation near the top plate where the system had a large thermal gradient \cite{AGL09} and where the top temperature was below $T_{\phi}$ while most of the sample remained in the vapor phase at temperatures above $T_{\phi}$. As $T_t$ was gradually lowered below $T_\phi$,  $\lambda_{eff}$ increased linearly above the single-phase value. In this regime $\lambda_{eff}$ was completely reproducible and independent of history. Shadowgraph images showed that condensation was by droplet formation rather than by film condensation.\cite{Co72,Dh98}  Even though $H$ vanishes at the CP, the largest enhancement $\delta \lambda_{eff}^*(P)$ of $\delta \lambda_{eff}(P,T)$ at a given pressure   {\it increased} rather dramatically as $P$ approached the critical pressure $P_{CP}$. Since the heat transport depends not only on $H$ but also on the rate of droplet formation, this result implies a droplet-nucleation rate that increased more rapidly than $H$ decreased as $P \rightarrow P_{CP}$. This result is qualitatively consistent with classical nucleation theory \cite{GMS83}; but clearly our system, with a large thermal gradient just below the top plate and with vigorous fluctuations, is more complicated than those treated before by that  or more advanced theories. Interestingly, $\delta \lambda_{eff}$ was essentially the same for two different samples, one with a finely machined copper top plate and the other with an optically flat sapphire top plate, suggesting that surface roughness did not influence the nucleation rate significantly and that the nucleation process was homogeneous.  

\begin{figure}
\includegraphics[width=2.3in]{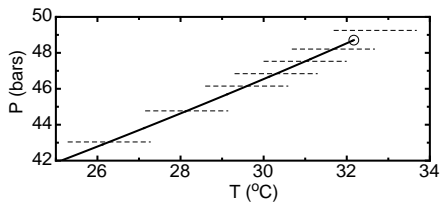}
\caption{Pressure-temperature plane showing the coexistence curve $T_{\phi}(P)$ (solid line) and the CP (circle) of C$_2$H$_6$. Dashed lines: some of the isobars used here.}
\label{fig:phasedia}
\end{figure}

The apparatus had been used for several previous investigations of turbulent RBC \cite{XBA00,AX01,NBFA05,FBNA05,BFNA05}. Sequentially we installed two different high-pressure sample cells. The first, cell A,  was a cylinder with $L$=7.62 cm and $D$=7.63cm. It had been used for turbulent heat-transport measurements in gases \cite{AFFGL07,ACFFGLS08}. Its top and bottom consisted of thick copper plates, with finely machined inner surfaces, that fit closely into a side wall made of high-tensile-strength stainless steel.  For flow-visualization with the shadowgraph method \cite{BBMTHCA96} we used cell B which had $D = 10.16$ cm and $L = 9.84$ cm. It had an optically flat sapphire top plate but a copper bottom plate with a polished surface and an evaporated gold film that served as a mirror. The measurements of $\lambda_{eff}$  reported here are for cell A, but cell B yielded largely equivalent results.  Heat was applied at the sample bottom by a metal-film heater covering the bottom-plate area uniformly.  The top plate was cooled by a circulating water bath.  Both $T_b$ and $T_t$ were held constant within a milli-Kelvin or better. 
The sample was connected to an external volume through a capillary. The temperature of this volume was controlled in a feedback loop with a pressure gage so as to hold the sample pressure constant within $10^{-3}$ bars.  C$_2$H$_6$ has well known properties \cite{FIE91, LTS92} and a conveniently located CP at $ T_{CP} = 32.172^\circ$C, $P_{CP} = 48.72$ bars.  Its phase-separation curve $T_{\phi}(P)$ is shown in Fig.~\ref{fig:phasedia}.

\begin{figure}
\includegraphics[width=2.3in]{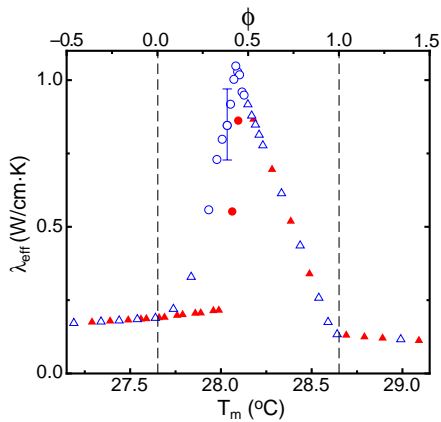}
\caption{Effective conductivity $\lambda_{eff}$ as a function of the mean fluid temperature  $T_m$ (lower abscissa)  and the temperature fraction $\phi=0.5+(T_m - T_{\phi}(P))/\Delta T$ (upper abscissa) for $P=44.77$ bars and $\Delta T=1.00$ K. Triangles:  $\lambda_{eff}$ is time-independent. Circles: $\lambda_{eff}$ is time-dependent (the time averaged values are shown; typical fluctuation amplitudes are indicated by the bar at $T_m=28.03^\circ$C). Solid (open) symbols:  $T_m$ was increased (decreased).}
\label{fig:lam_eff}
\end{figure}

In Fig.~\ref{fig:lam_eff} measurements of $\lambda_{eff} = QL/ \Delta T$ ($Q$ is the heat-current density) at constant  $P=44.77$ bars and $\Delta T=1.00$ K are plotted as a function of $T_m$ (lower abscissa) or of the temperature fraction $\phi=0.5+(T_m - T_{\phi}(P))/\Delta T$ (upper abscissa). For this pressure  $T_{\phi}(P)=28.148^\circ$ C. The two vertical dashed lines show the temperatures $T_m=T_{\phi}(P) \pm \Delta T/2$.
For $\phi > 1$ the entire sample was in the vapor phase and $\lambda_{eff}$ was close to values obtained from an extrapolation of measurements made before \cite{AFFGL07,ACFFGLS08} but further away from $T_\phi$. For $\phi < 0$ the entire sample was in the liquid phase, and again  $\lambda_{eff}$ was consistent with other measurements. As $\phi$ was lowered from $\phi > 1$ into the two-phase region below $\phi = 1$,  $\lambda_{eff}$ initially increased linearly as a function of $\phi$ and reached a maximum value $\lambda^*_{eff}$ at $\phi^* \simeq 0.43$ that was nearly an order of magnitude larger than in the vapor.
There was a sharp onset of this heat-transport enhancement, but there was no discontinuity. The solid (open) symbols correspond to data taken with increasing (decreasing) $T_m$ or $\phi$. One sees that  $\lambda_{eff}$ was independent of this past history and highly reproducible. This region of linear increase, we shall show, corresponded to a sample filled with vapor but with droplet condensation occurring at the top plate. As $\phi$ dropped slightly below $\phi^*$, the heat transport became time dependent and the time-averaged values became history dependent. Further reduction of 
 $\phi$ (but still with $\phi > 0$) led once more to time-independent states, but with  relatively small heat-transfer enhancements that also varied from run to run. This parameter range corresponded to a sample filled with liquid and with vaporization (or ``boiling") occurring at the bottom plate\cite{OVLP08}. The transition from a mostly vapor-filled to a mostly liquid-filled state near $\phi = \phi^*$ could easily be seen in the experiment because it led to a discontinuous increase of the temperature of the external volume used for the pressure regulation.  

\begin{figure}
\includegraphics[width=2.3in]{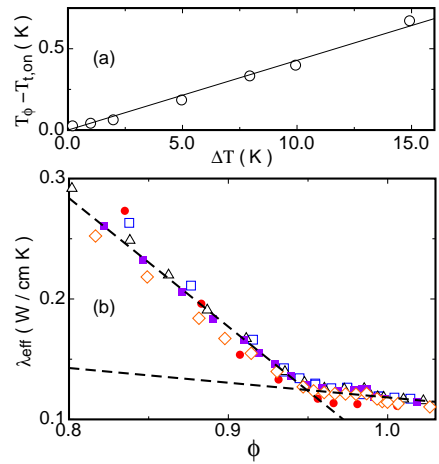}
\caption{(a): The shift $T_\phi - T_{t,on}$ of the onset of heat-current enhancement as a function of the applied temperature difference $\Delta T$. The solid line has a slope of 0.043. (b): $\lambda_{eff}$ as a function of $\phi$. Solid circles: $\Delta T = 2.00$K. Open squares: $\Delta T=5.00$K. Triangles: $\Delta T = 8$K. Solid squares: $\Delta T= 10.00$K. Open diamonds: $\Delta T= 15.00$K.  The pressure was 43.04 bars.}
\label{fig:onset2}
\end{figure}

An interesting aspect of the onset of the heat-transport enhancement is that it occurred at $T_{t,on} < T_{\phi}(P)$. The shift $T_{\phi} - T_{t,on}$ increased roughly linearly with $\Delta T$, as shown in Fig.~\ref{fig:onset2}a for a pressure of 43.04 bars. This linear increase implies a constant shift of the temperature fraction $\phi_{on}$ at onset below $\phi = 1$; it is illustrated in  Fig.~\ref{fig:onset2}b where $1 - \phi_{on} \simeq 0.05$ independent of $\Delta T$.

\begin{figure}
\includegraphics[width=2.8in]{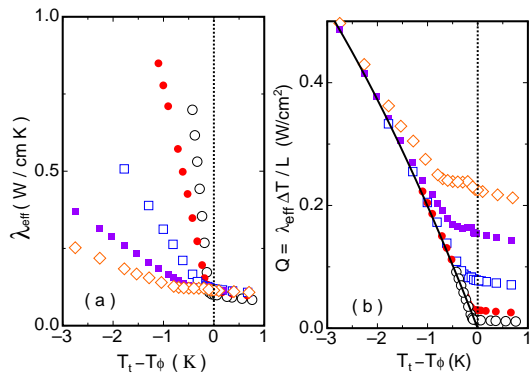}
\caption{(a): The effective conductivity $\lambda_{eff}$ and (b): $Q$ at a pressure of 43.04 bars and at several applied temperature differences $\Delta T$. Open circles: $\Delta T = 1.00$K. Solid circles: $\Delta T = 2.00$K. Open squares: $\Delta T=5.00$K. Solid squares: $\Delta T= 10.00$K. Open diamonds: $\Delta T= 15.00$K. Solid line in (b): $Q = -0.212(T_t-T_\phi) - 0.013(T_t-T_\phi)^2$ W/cm$^2$.}
\label{fig:onset1}
\end{figure}

In Fig.~\ref{fig:onset1}a one sees that a larger $\Delta T$ leads to a slower increase of $\lambda_{eff}$ with decreasing $T_t$. Figure~\ref{fig:onset1}b reveals that, for sufficiently large $T_\phi - T_t $, the data for the heat-current density $Q$ approach a single curve, showing that the heat transport is determined by $T_\phi - T_t$ and not by $\Delta T$.

\begin{figure}
\includegraphics[width=2.75in]{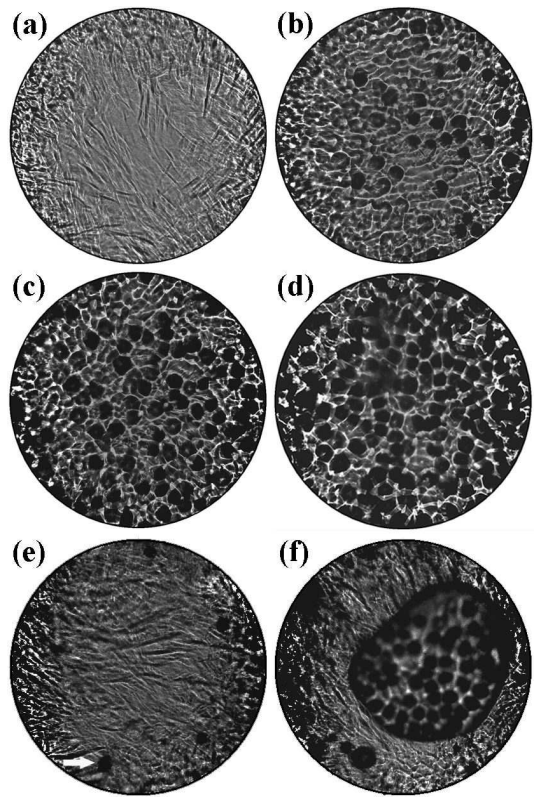}
\caption{Shadowgraph images from cell B at different mean temperatures with (a) $\phi=1.01$, (b) 0.92, (c) 0.85, (d) 0.71, (e) 0.32, and (f) 0.58. For this experiment $P=41.37$ bars, $\Delta T= 0.50K$, and $T_m$ was increasing. Movies for $\phi = 1.01$ 0.92, 0.32, and 0.58 are available elsewhere \cite{commentMO09}.}
\label{fig:shadow}
\end{figure}

In Fig.~\ref{fig:shadow} we show shadowgraph images taken in cell B. It is difficult to interpret these images quantitatively because they give a vertical average and the vertical location of any feature remains unresolved. Nonetheless they give useful qualitative information. Image (a) for $\phi = 1.01$ is for the single-phase vapor region.  At this point the Rayleigh number $Ra = \beta g \Delta T L^3/(\kappa\nu)$ ($g$ is the gravitational acceleration and $\beta,~\nu$, and $\kappa$ are the thermal expansion coefficient, kinematic viscosity, and thermal diffusivity respectively) was about $3.8\times 10^{10}$  and the Nusselt number $Nu = \lambda_{eff}/\lambda$ ($\lambda$ is the diffusive thermal conductivity) was close to 220. The structure seen in the shadowgraph corresponds to plume activity and fluctuations of the highly turbulent single-phase system. A movie for this case \cite{commentMO09} shows the turbulent time dependence and reveals the existence of a large-scale circulation which swept the structures along \cite{FA04,FBA08}. Image (b), for $\phi = 0.92$,   is in the two-phase region where a thin layer of fluid in the top thermal boundary layer had been rendered meta-stable. One sees a new feature: there were numerous small dark circles which we interpret to be liquid droplets. They moved laterally and were swept toward the side wall, we believe by the prevailing large-scale circulation. A movie of this case can also be found elsewhere \cite{commentMO09}. Images (c) and (d) are for $\phi = 0.85 $ and 0.71 respectively, and the abundance of droplets is seen to have increased as $\phi$ decreased.

Images (e) and (f) are for $0 < \phi < \phi^*$ where the sample is mostly liquid-filled. Somewhat above $\phi = 0$ boiling started but $\lambda_{eff}$ was not enhanced  very much. A small number of isolated gas bubbles can be seen in (e), with one of them identified by the small white arrow in the lower left part. The bubbles meander chaotically in the lateral direction and, upon rising, re-dissolve in the cooler sample interior. At larger $\phi < \phi^*$, the small bubbles collect in one large bubble located under the top plate as shown in (f). Inside that bubble condensation is taking place as evident from the many drops that form within it. These processes are illustrated better by the movies \cite{commentMO09} for (5e) and (5f). In the boiling range, with $\phi < \phi^*$, measurements of $\lambda_{eff}$ were irreproducible from one run to another. 

\begin{figure}
\includegraphics[width=2.3in]{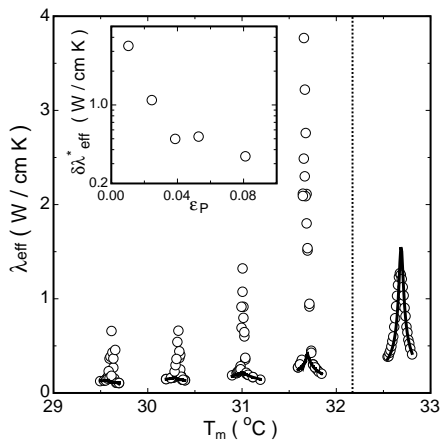}
\caption{The effective conductivity $\lambda_{eff}$, measured with $\Delta T = 0.1$ K,  as a function of $T_m$ for different isobars. From left to right, the data are for $P$=46.14, 46.83,47.52, 48.21, and 49.25 bars. The solid curves are estimates for single-phase turbulent convection using the properties at $T_m$ and the predictions of Ref.~\cite{GL01}. The dotted line shows the critical temperature. The inset shows the maximum excess contribution $\delta \lambda^*_{eff} = \lambda^*_{eff} - \lambda^*_{RB}$  from the nucleation process  to $\lambda_{eff}$ as a function of the reduced pressure $\epsilon_p$.}
\label{fig:lameff2}
\end{figure} 

The latent heat vanishes at the CP, and above the CP the liquid and gas states become indistinguishable. Thus one might expect initially that the heat-transport enhancement in the two-phase region should be weakened as the CP is approached and cease to exist for $P > P_{CP}$. In Fig.~\ref{fig:lameff2} we show measurements of $\lambda_{eff}$ with $\Delta T=0.1$K at different $P$. Remarkably, as $P \rightarrow P_{CP}$,  $\lambda^*_{eff}$ increased dramatically. As $P$ exceeded $P_{CP}$, $\lambda^*_{eff}$ decreased again. 

At the CP $\beta$ diverges and  $\kappa$ vanishes. As a result $Ra$ and the Prandtl number $Pr = \nu/\kappa$ become infinite. For these rapidly varying conditions we estimated the heat transport contributed by single-phase turbulent Rayleigh-B\'enard convection $\lambda_{RB}$ using the predictions from Ref.~\cite{GL01} with fluid properties evaluated at  $T_m$. These estimates are shown by the solid curves in Fig.~\ref{fig:lameff2}. For $P$ above $P_{CP}$ there is reasonable agreement between the prediction and the measurements. Below $P_{CP}$, as expected, the estimate is much too small and there is a much larger contribution from condensation or boiling.   
The inset of Fig.~\ref{fig:lameff2} shows the excess $\delta \lambda^*_{eff}=(\lambda^*_{eff}-\lambda^*_{RB})$ as a function of the reduced pressure $\epsilon_p=(P_{CP} - P)/P_{CP}$ ($\lambda^*_{RB}$ is the maximum value of $\lambda_{RB}$ at a given $P$). The maximum heat transport contributed by the nucleation process increased by an order of magnitude as the fluid pressure approached $P_{CP}$.

In this Letter we reported on heat-transport measurements by turbulent Rayleigh-B\'enard convection of ethane under conditions where the applied temperature difference $\Delta T$ straddled the liquid-vapor coexistence curve $T_\phi(P)$. As the top temperature $T_t$ was lowered quasi-statically below $T_\phi$, the effective conductivity $\lambda_{eff}$ was enhanced by droplet condensation at the sample top. The droplet formation was observed by shadowgraphy. With decreasing $T_t$ and starting at $T_{t,on} < T_\phi$, $\lambda_{eff}$ initially increased linearly and reached a maximum $\lambda^*_{eff}$ at $T^*_t$ that was an order of magnitude or more larger than typical values in the single-phase regions. Here $\lambda_{eff}$ was reproducible and history independent. The shift of the onset $T_\phi - T_{t,on}$ increased roughly linearly with  $\Delta T$.  

The maximum enhancement $\delta \lambda^*_{eff}$ above the single-phase value increased dramatically as the pressure approached the critical value $P_{CP}$. Since the latent heat $H$ vanishes at the critical point, this implies that the droplet-nucleation rate increased at a sufficiently large rate to more than overcome the diminished contribution from $H$. We do not know of a droplet-nucleation theory that would be applicable quantitatively in the presence of the steep thermal gradient and the vigorous fluctuations characteristic of turbulent convection. However, under the more benign circumstances of an isothermal meta-stable fluid classical nucleation theory \cite{GMS83} suggests that the nucleation rate should be proportional to $\exp(-\Delta F/k_BT)$, and that the difference in free energy $\Delta F$ between the vapor and the droplet should vanish at the critical point. Qualitatively this implies an enhanced nucleation rate as $P \rightarrow P_{CP}$.

This work was supported by the National Science Foundation through Grant DMR07-02111.


\end{document}